\documentclass[conference]{./IEEEtran}

\IEEEoverridecommandlockouts

\ifCLASSINFOpdf
\else
\fi
  \usepackage{tipa}
  \usepackage{graphicx}
	\usepackage{epsfig}
	\usepackage{epstopdf}
	\usepackage{upgreek}
	\usepackage[nolist,nohyperlinks]{acronym}

\usepackage[cmex10]{amsmath}

\usepackage{color, soul}

\usepackage{subfigure}

\usepackage{cite}

\begin{acronym}
\acro{NoC}{Network-on-Chip}
\acro{WNoC}{Wireless Network-on-Chip}
\acro{EM}{Electromagnetic}
\acro{SiO$_2$}{Silicon Dioxide}
\acro{AIN}{Aluminum nitride}
\acro{SiP}{System-in-Package}
\acro{SDM}{Software-defined metamaterial}
\acro{RZ}{Return-to-Zero}
\acro{MAC}{Medium Access Control}
\acro{THz}{Terahertz}
\acro{RC}{Resistive-Capacitive}
\acro{ENOB}{Effective Number of Bits}
\acro{mm-Wave}{millimeter-wave}
\acro{OOK}{On-Off Keying}
\acro{PLL}{Phase-Locked Loop}
\acro{DAC}{Digital-to-Analog Converter}
\acro{ADC}{Analog-to-Digital Converter}
\acro{SAR}{Successive Approximation Register}
\acro{CDMA}{Code Division Multiple Access}
\acro{OFDM}{Orthogonal Frequency Division Multiplexing}
\end{acronym}

\begin{document}

% ---------------------------------------------------
% Title
% ---------------------------------------------------

% Titles are generally capitalized except for words such as a, an, and, as,
% at, but, by, for, in, nor, of, on, or, the, to and up, which are usually
% not capitalized unless they are the first or last word of the title.
% Linebreaks \\ can be used within to get better formatting as desired.
% Do not put math or special symbols in the title.

\title{Data Conversion in Area-Constrained Applications: the Wireless Network-on-Chip Case
\thanks{This work has been supported in part by the European Commission under grant H2020-FETOPEN-736876 (VISORSURF) and the Spanish MINECO under contract TEC2017-90034-C2-1-R (ALLIANCE).} %, and the Catalan Institution for Research and Advanced Studies (ICREA).}
} %% FIX: Icrea if albert

%
%A Feasibility Analysis for Analog-Digital Conversion in Wireless Network-on-Chip

% ---------------------------------------------------
% Authors, Affiliation, and Acknowledgment
% ---------------------------------------------------

% use a multiple column layout for up to three different
% affiliations
%\author{
%\IEEEauthorblockN{Sergi Abadal, Alejandro L\'{o}pez-Lao, Eduard Alarc\'{o}n}
%\IEEEauthorblockA{NaNoNetworking Center in Catalunya (N3Cat), Universitat Polit\`{e}cnica de Catalunya (UPC), Barcelona, Spain\\%
%Corresponding Email: abadal@ac.upc.edu%
%}
%}

\author{
\IEEEauthorblockN{Sergi Abadal}
\IEEEauthorblockA{\textit{Computer Architecture Department}\\
\textit{Universitat Polit\`{e}cnica de Catalunya}\\
Barcelona, Spain\\
abadal@ac.upc.edu}
%\and
%\IEEEauthorblockN{Alejandro L\'{o}pez-Lao}
%\IEEEauthorblockA{\textit{Computer Architecture Department}\\
%\textit{Universitat Polit\`{e}cnica de Catalunya}\\
%Barcelona, Spain\\
%alejandrolopezlao@gmail.com}
\and
\IEEEauthorblockN{Eduard Alarc\'{o}n}
\IEEEauthorblockA{\textit{Electrical Engineering Department}\\
\textit{Universitat Polit\`{e}cnica de Catalunya}\\
Barcelona, Spain\\
eduard.alarcon@upc.edu}
}
%\IEEEauthorblockA{
%\IEEEauthorrefmark{2}Electrical Engineering Department\\
%Iran University of Science and Technology (IUST), Tehran, Iran}

% conference papers do not typically use \thanks and this command
% is locked out in conference mode. If really needed, such as for
% the acknowledgment of grants, issue a \IEEEoverridecommandlockouts
% after \documentclass

% for over three affiliations, or if they all won't fit within the width
% of the page, use this alternative format:
% 
%\author{\IEEEauthorblockN{Michael Shell\IEEEauthorrefmark{1},
%Homer Simpson\IEEEauthorrefmark{2},
%James Kirk\IEEEauthorrefmark{3}, 
%Montgomery Scott\IEEEauthorrefmark{3} and
%Eldon Tyrell\IEEEauthorrefmark{4}}
%\IEEEauthorblockA{\IEEEauthorrefmark{1}School of Electrical and Computer Engineering\\
%Georgia Institute of Technology,
%Atlanta, Georgia 30332--0250\\ Email: see http://www.michaelshell.org/contact.html}
%\IEEEauthorblockA{\IEEEauthorrefmark{2}Twentieth Century Fox, Springfield, USA\\
%Email: homer@thesimpsons.com}
%\IEEEauthorblockA{\IEEEauthorrefmark{3}Starfleet Academy, San Francisco, California 96678-2391\\
%Telephone: (800) 555--1212, Fax: (888) 555--1212}
%\IEEEauthorblockA{\IEEEauthorrefmark{4}Tyrell Inc., 123 Replicant Street, Los Angeles, California 90210--4321}}

% use for special paper notices
%\IEEEspecialpapernotice{(Invited Paper)}

% make the title area
\maketitle

% ---------------------------------------------------
% Abstract
% ---------------------------------------------------

% As a general rule, do not put math, special symbols or citations
% in the abstract
\begin{abstract} 
Network-on-Chip (NoC) is currently the paradigm of choice to interconnect the different components of System-on-Chips (SoCs) or Chip Multiprocessors (CMPs). As the levels of integration continue to grow, however, current NoCs face significant scalability limitations and have prompted research in novel interconnect technologies. Among these, wireless intra-chip communications have been under intense scrutiny due to their low latency broadcast and architectural flexibility. Thus far, the practicality of the idea has been studied from the RF front-end and the network interface perspectives, whereas little to no attention has been placed on another essential component: the data converters. This article aims to fill this gap by providing a comprehensive analysis of the requirements of the scenario, as well as of the current performance and cost trends of \acp{ADC}. Based on Murmann's data, we demonstrate that \acp{ADC} will not be a roadblock for the realization of wireless intra-chip communications although current designs do not meet their demands fully.
\end{abstract}

% ---------------------------------------------------
% Keywords
% ---------------------------------------------------

\begin{IEEEkeywords}
Wireless Network-on-Chip; High-Speed Data Conversion; DAC; ADC
\end{IEEEkeywords}

% For peer review papers, you can put extra information on the cover
% page as needed:
% \ifCLASSOPTIONpeerreview
% \begin{center} \bfseries EDICS Category: 3-BBND \end{center}
% \fi
%
% For peerreview papers, this IEEEtran command inserts a page break and
% creates the second title. It will be ignored for other modes.
\IEEEpeerreviewmaketitle

% FIX: if we even want to extend this to a journal, we should
% FIX: 1) better define the system model, methodology and metrics.
% FIX: 2) provide a minimum background on the types of ADCs.
% FIX: 3) be more precise on the number of bits that we need for OOK, PAM, QAM, and crazy OFDM or CDMA ~ is CDMA only requiring higher sampling speed? Maybe some mathematics.
% FIX: 4) extend the discussion to the types of ADC's as well. This means
% FIX: --- repeat figures 4-6, indicating the type.
% FIX: --- cite the trends from other papers.
% FIX: --- maybe try to replicate what Jonsson did.

% ---------------------------------------------------
% Introduction
% ---------------------------------------------------

\acresetall

\section{Introduction} \label{sec:introduction}
% Intro NoC 
Network-on-Chip (NoC) has become the paradigm of choice to interconnect cores and memory within a chip. However, recent years have seen a significant increase in the core density and, within this context, it becomes increasingly difficult to meet the on-chip communication requirements with conventional NoCs alone \cite{Bertozzi2014}. Their limited scalability is gradually turning communication, not computation, into the performance bottleneck in parallel processing. New solutions are thus required to avoid slowing down progress in the manycore era \cite{Kim2012Survey}.

% Intro WNoC 
Advances in integrated \ac{mm-Wave} antennas \cite{Markish2015, Cheema2013} and transceivers \cite{Laha2015, Subramaniam2017} have led to the proposal of \ac{WNoC} as a potential alternative to conventional NoC fabrics \cite{Matolak2012}. In a \ac{WNoC}, a set of cores is augmented with transceivers and antennas capable of modulating and radiating the information. RF signals propagate through the computing package and can be demodulated by all tuned-in receivers. The main advantage of this approach is that distant cores can communicate with low latency as propagation occurs nearly at the speed of light. In fact, communication is naturally broadcast. Further, the wireless approach provides an architectural flexibility very hard to achieve with wired alternatives.

\begin{figure}[!t]
\centering
\includegraphics[width=\columnwidth]{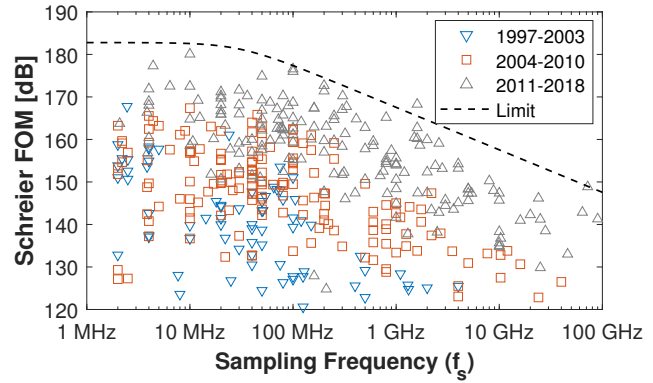}
\vspace{-0.5cm}
\caption{Evolution of the ADC performance--efficiency over the last two decades. The Schreier's figure of merit, as defined in Equation \eqref{eq:FOM}, is a metric of efficiency and is represented as a function of the sampling frequency.}
\label{fig:timeX}
\vspace{-0.1cm}
\end{figure}

% Motivation --- lots of works, but generally no consideration for the a priori expensive ADC/DAC requirements
Due to its potential, \acp{WNoC} have been investigated extensively from different standpoints \cite{Matolak2012, Kim2015, Abadal2018}. At the circuit level, most efforts have focused on the design of the analog front end \cite{Subramaniam2017} or aspects around it, e.g., power gating \cite{Mondal2017}. In contrast, and despite being a critical function in any wireless system, data conversion has been generally overlooked or taken for granted. However, at speeds over 10 Gb/s and given the evident resource limitations of nowadays chips, such an assumption is far from trivial even for simple modulations. Advanced options such as \ac{CDMA} \cite{Vijayakumaran2014} or \ac{OFDM} \cite{Gade2017} may be simply out of question in this scenario. 

% Contribution
In this paper, we aim to provide a feasibility analysis of data conversion in the \ac{WNoC} scenario. We first estimate the data conversion demands of the \ac{WNoC} paradigm by the order of magnitude. Then, we use historical figures from actual converter implementations \cite{Murmann2013, Murmann2015} to update existing performance and energy efficiency predictions, as illustrated in Figure \ref{fig:timeX}. Area scaling trends, which have been given much less attention in the literature \cite{Jonsson2011, Verhelst2012}, are also studied in depth here in light of the chip real state restrictions of the on-chip scenario. With this, we are able to assess whether data conversion will be a roadblock in the realization of the \ac{WNoC} or not, and under which conditions. 

In a broader sense, the analysis would be applicable to other wireless applications with evident resource constraints such as Wireless Nanosensor Networks (WNSNs) \cite{Akyildiz2010} or Software-Defined Metamaterials (SDMs) \cite{Tasolamprou2018}. In any case, and to the best of the author's knowledge, this is the first gap analysis relative to data conversion in emerging area-constrained applications.

%we cannot consider \ac{WNoC} as the only wireless communications scenario with evident area and energy constraints. Other applications such as Wireless Nanosensor Networks (WNSNs) \cite{} or Software-Defined Metamaterials (SDMs) \cite{} will require reasonably fast but ultra-efficient converters.

% Reminder
The remainder of this paper is organized as follows. Section \ref{sec:considerations} reviews the main wireless channel and physical considerations of the \ac{WNoC} case, to then derive a rough quantification of the requirements for data converters. Section \ref{sec:trend} discusses how close are current designs from being able to accommodate the predicted \ac{WNoC} requirements, whereas Section \ref{sec:discussion} extrapolates future behavior from on-going performance and cost trends. Finally, Section \ref{sec:conclusion} concludes the paper.

\section{Wireless Network-on-Chip: System Considerations}
\label{sec:considerations}
\begin{figure}[!t]
\centering
\includegraphics[width=\columnwidth]{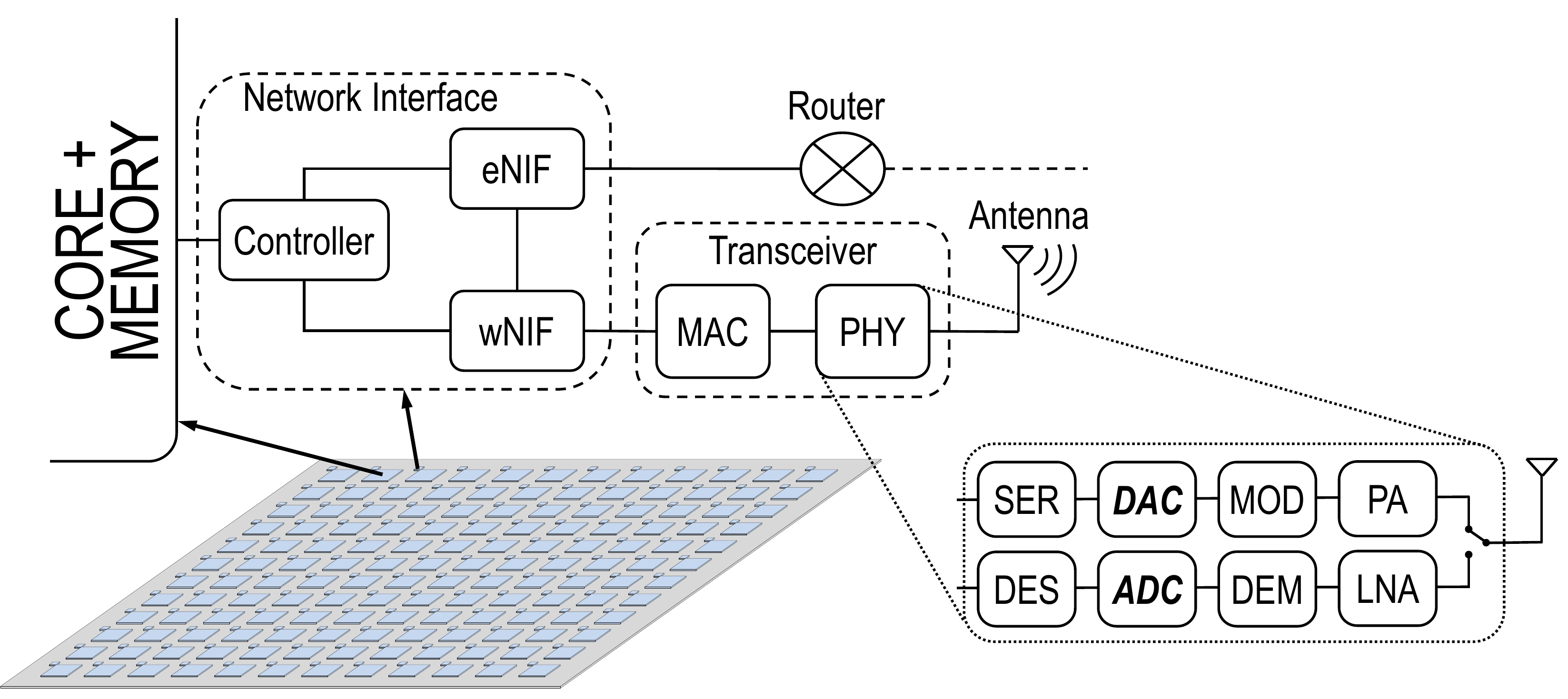}
\vspace{-0.4cm}
\caption{Schematic of a \ac{WNoC} architecture with wireless interfaces at the core level. At the transceiver, a generic physical layer module is considered. At the uplink, the transceiver contains a serializer (SER), digital-to-analog converter (DAC), modulator (MOD) and power amplifier (PA). At the downlink, it includes a low-noise amplifier (LNA), demodulator (DEM), analog-to-digital converter (ADC) and deserializer (DES).}
\label{fig:schematic}
\vspace{-0.1cm}
\end{figure}

Figure~\ref{fig:schematic} exemplifies the \ac{WNoC} paradigm, basically comprising a co-integration of antennas and transceivers with cores in complement of the wired \ac{NoC}. As an asymptotic case, let us assume that wireless communication capabilities are given at the core level. However, note that other works may assume a reduced number of wireless interfaces through clustering or co-integration with the routers \cite{Sujay2012}. In either case, \acp{WNoC} are only considered in large manycores (i.e., tens or hundreds of cores), where communications can become a performance bottleneck.

In this context, the network interface performs address translation, load balancing, and admission control; whereas the \ac{MAC} module performs the usual actions to ensure that all nodes can access the shared medium without collisions. At the physical layer, four functions common to any wireless system interfaced to a digital architecture are performed: serialization, data conversion, modulation, and power amplification. Next, we describe several design considerations that have a potential impact on the data conversion process.

%\begin{itemize}
%\item\textbf{Serialization:} as information comes from the processor core and cache hierarchy, control and data are bundled and need to be serialized for transmission.  
%\item\textbf{Data Conversion:} as in any wireless system interfaced to a digital source, data must be converted to analog signals and \emph{vice versa}. This is the crux of the paper.
%\item\textbf{Modulation:} 
%\item\textbf{Power amplification:}
%\end{itemize}
\noindent
\textbf{General Considerations.} Table \ref{tab:reqs} provides a rough quantification of the communication requirements in a \ac{WNoC}, substantiated by the following. As mentioned above, communications are crucial in manycores as they can become the system bottleneck if not served well. For this reason, latency and throughput objectives are set to very ambitious levels for wireless communications, with latencies in the nanosecond scale and throughputs in the order of tens of Gb/s, to compete with chip-wide wired \ac{NoC} options. Additionally, the error rates are generally assumed to be similar to that of \ac{RC} wires. 

\begin{table}[!t] 
%\caption{Manycore Communication Requirements}
\caption{Wireless Manycore Scenario Requirements}
\label{tab:reqs}
\footnotesize
\centering
\begin{tabular}{ll} 
\hline
{\bf Metric} & {\bf Value} \\
\hline
Transmission Range & 0.1--10 cm \\
Node Density & 10--1000 nodes/cm\textsuperscript{2} \\
Network Throughput & 10--100 Gb/s \\
Latency & 1--100 ns \\
Bit Error Rate (BER) & 10\textsuperscript{-15} \\
Transceiver Energy & 1--10 pJ/bit \\
Transceiver Area & 0.01--1 mm\textsuperscript{2} \\
\hline
\end{tabular}
\vspace{-0.3cm}
\end{table}

Besides high performance, \acp{WNoC} must also seek cost efficiency both in terms of area and power. Area constraints are evident given that the dimensions of a chip, typically 20$\times$20 mm\textsuperscript{2}, do not scale up with the number of cores \cite{HuangMICRO2011}. Multiprocessor systems are also energy-aware, if not energy-limited, because the available total power does not scale with the number of cores either, mainly due to heat dissipation issues \cite{HuangMICRO2011}. Therefore, manycore systems account for a strict power budget so that the Thermal Design Point (TDP), varying from tens to a few hundreds of Watts, is always respected. 

Assuming a 100-core processor in a 450 mm\textsuperscript{2} chip with the TDP of a Xeon Phi (210 W), we will thus have that each core can only take 4.5 mm\textsuperscript{2} and at most 2.1 W of sustained power including the processor, memory, and communication sub-systems. Optimistically assuming the same budget for the three sub-systems, the \ac{NoC} (including the wireless part, if any) should not exceed 1.5 mm\textsuperscript{2} and not take more of 700 mW per core. Assuming again an equitable distribution of resources and neglecting network interface and MAC overheads, we would estimate the \ac{WNoC} to have a budget of around 0.75 mm\textsuperscript{2} and 350 mW per core (3.5 pJ/bit at 100 Gb/s or around 35 pJ/bit at 10 Gb/s). Let this estimation serve as reasonable limits for the cost of a \ac{WNoC}, noting that they would be increased or reduced depending on the actual distribution of resources and the number of cores. 

%As a consequence, \acp{NoC} can take 
%The importance of communications in a manycore are also reflected 
%Another consequence of the importance of internal communications in a manycore is that the 
\begin{figure*}[!t]
\centering
\includegraphics[width=0.48\textwidth]{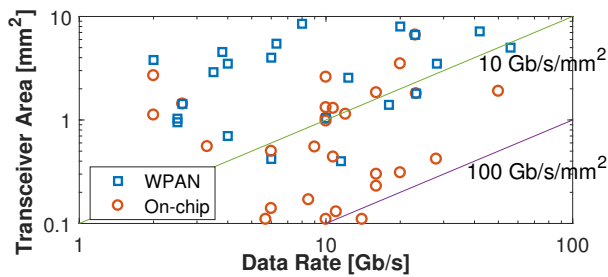}
\includegraphics[width=0.48\textwidth]{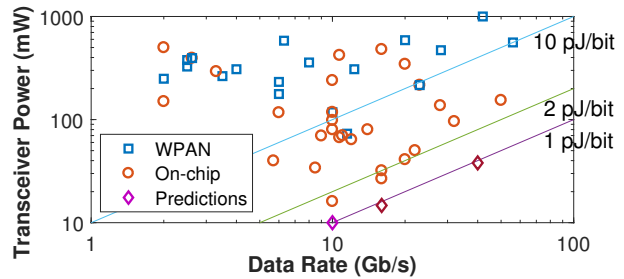}
\vspace{-0.2cm}
\caption{Area and power of recent transceiver designs for Wireless Personal Area Networks (WPAN) and chip-scale wireless communications, represented as functions of the bitrate. Data extracted from \cite{AbadalTON, Tasolamprou2018} and references therein.}
\label{fig:transc}
\vspace{-0.1cm}
\end{figure*}

\noindent
\textbf{Transceiver cost.} Most of the overhead of the \ac{WNoC} is expected to come from the analog front-end at the physical layer. To estimate it, one can take base on existing designs optimized for this scenario \cite{Subramaniam2017}, which report performance and costs compliant with the above estimations. Another route is to include these designs in a wider exploration of the state of the art, which may allow us to also obtain trends. In this respect, Figure \ref{fig:transc} shows the area and power of a set of transceiver designs for multi-Gb/s short-range wireless communications reported in the period of 2010--2018. Note that the antenna is not included in such analysis.

The main outcome of Figure \ref{fig:transc} is the confirmation of that (1) the throughput objectives are well achievable, and (2) most of the area and power budgets for \ac{WNoC} will be taken by the transceiver and the antenna. Therefore, there is not much room to spare for the data converters and the serializer circuits. For the purpose of this article, let us assume that at most 10\% of the whole transceiver area and power will be devoted to data conversion. In other words:
\begin{itemize}
\item Conversion circuits should occupy less than $\sim$0.1 mm\textsuperscript{2}. 
\item Conversion should consume less than $\sim$1 pJ/bit.
\end{itemize}

\noindent
\textbf{Modulation.} \ac{WNoC} uses \ac{mm-Wave} frequencies and points to the \ac{THz} band so that the antennas become commensurate with the cores in manycore settings. This pushes the requirements of the components of the transceiver limiting, together with the area and power constraints themselves, the complexity of the underlying modulation. As a result, most works in \ac{WNoC} assume simple modulation schemes such as \ac{OOK} and non-coherent detection \cite{AbadalTON}. Modulations requiring detection or phase or precise synchronization are avoided whenever possible, as \ac{PLL} circuits are extremely power-hungry. Simplicity, together with the stringent error rate requirements, are also the main reasons of advocating for modulations with low spectral efficiency. Equalization and other advanced signal processing methods are also out of question for the same reasons.

Simplicity in the modulation has several consequences at the data conversion stage, such as:
\begin{itemize}
\item Depending on the transmitting circuit topology and modulation, \acp{DAC} may be completely bypassed. 
\item At the \ac{ADC}, the sampling frequency will be pushed to speeds over tens of GS/s to comply with the throughput requirements. 
\item At the \ac{ADC}, the required \ac{ENOB} will be quite low as very few bits per sample are required (potentially down to one).
\end{itemize} 
%Current proposals for WNoC use simple modulations at around 60 GHz to minimize area and power [?], whereas future trends point to even higher Radio Frequency (RF) frequencies to further reduce area [?].

\noindent
\textbf{Wireless Channel.} Unlike in other wireless scenarios, communications in \ac{WNoC} take place in an enclosed and static environment \cite{Matolak2013CHANNEL}. This has different implications on the design of the physical and MAC layers of the protocol stack. For instance, the enclosed nature of the chip package leads to low path loss exponents \cite{Timoneda2018}, but is also expected to lead to long delay spreads. Fortunately, the static environment could allow the development of opportunistic solutions at the receiver, perhaps employing \ac{RZ} techniques or adaptive decision circuits. Being static, the chip environment would also allow the detection of collisions through unconventional approaches, such as the comparison of the received RF power with the source address of the packet \cite{Abadal2018a}.

The main consequences of the above considerations on the data conversion are the following:
\begin{itemize}
\item Moderate oversampling may be needed to meet the data rate requirements.
\item Additional bits per sample may be also used to improve performance.
\end{itemize}

\section{Gap Analysis} 
\label{sec:trend}
Table \ref{tab:adcs} shows a summary of the \ac{ADC} requirements in \ac{WNoC} derived from the order-of-magnitude estimations made in Section \ref{sec:considerations}. Here, we discuss current figures of state-of-the-art \acp{ADC} to assess whether current designs can meet expected \ac{WNoC} requirements. To this end, we take base on the widely recognized dataset by Murmann, which is considered an exhaustive and representative survey of \ac{ADC} designs for the last 20 years \cite{MurmannADC}. At the time of this writing, this dataset contains more than 500 entries coming from the flagship conferences in Solid-State Circuits (IEEE ISSCC) and Very Large Scale Integration Circuits (IEEE VLSIC). We analyze performance, area, and power consumption.

\begin{table}[!t] 
%\caption{Manycore Communication Requirements}
\caption{ADC Requirements in Wireless Network-on-Chip}
\label{tab:adcs}
\footnotesize
\centering
\begin{tabular}{ll} 
\hline
{\bf Metric} & {\bf Value} \\
\hline
Signal bandwidth & $\geq$ 10 GHz \\
Nyquist frequency & $\geq$ 20 GHz \\
Oversampling & Null or moderate \\
ENOB & $\leq$ 4 bits \\ % FIX: is this true?
Area & $\leq$ 0.1 mm\textsuperscript{2} \\
Energy & $\leq$ 1 pJ/bit \\
\hline
\end{tabular}
\vspace{-0.3cm}
\end{table}

\begin{figure*}[!t]
\centering
\subfigure[Performance (ENOB/BW)\label{fig:perf}]{\includegraphics[width=0.32\textwidth]{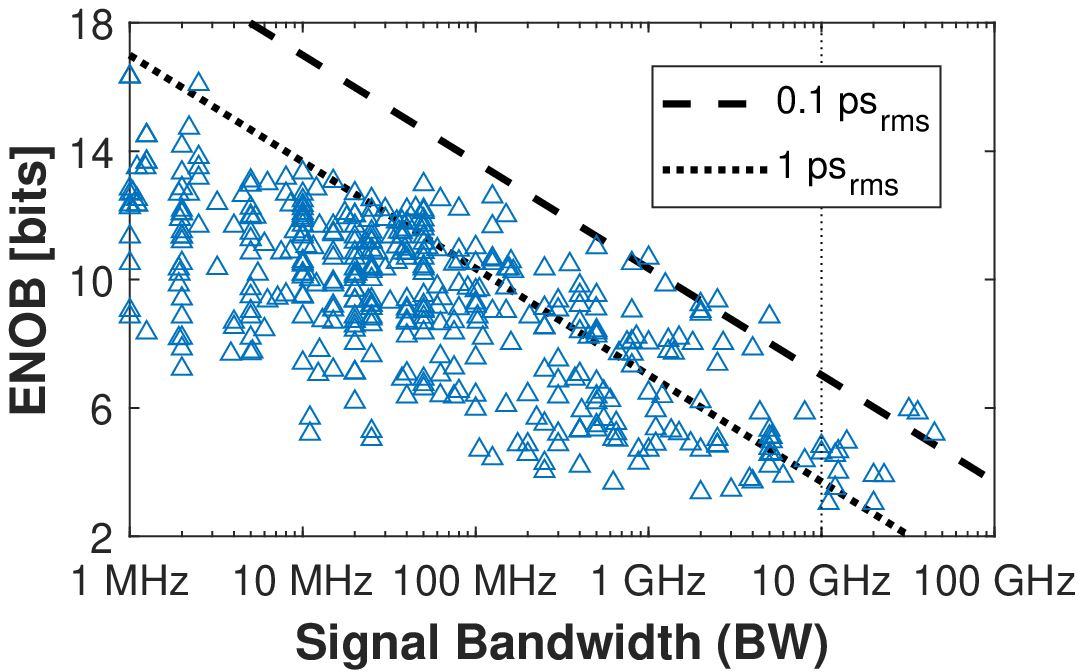}}
\subfigure[Energy ($E_{bit}$)\label{fig:power1}]{\includegraphics[width=0.32\textwidth]{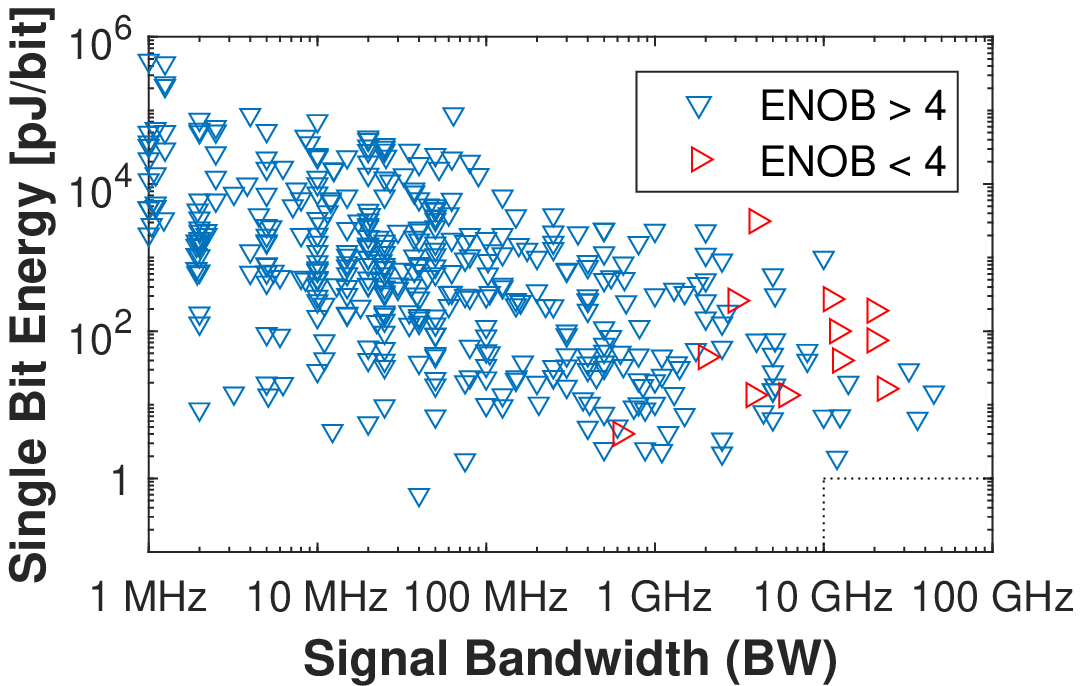}}
\subfigure[Area ($A$)\label{fig:area1}]{\includegraphics[width=0.32\textwidth]{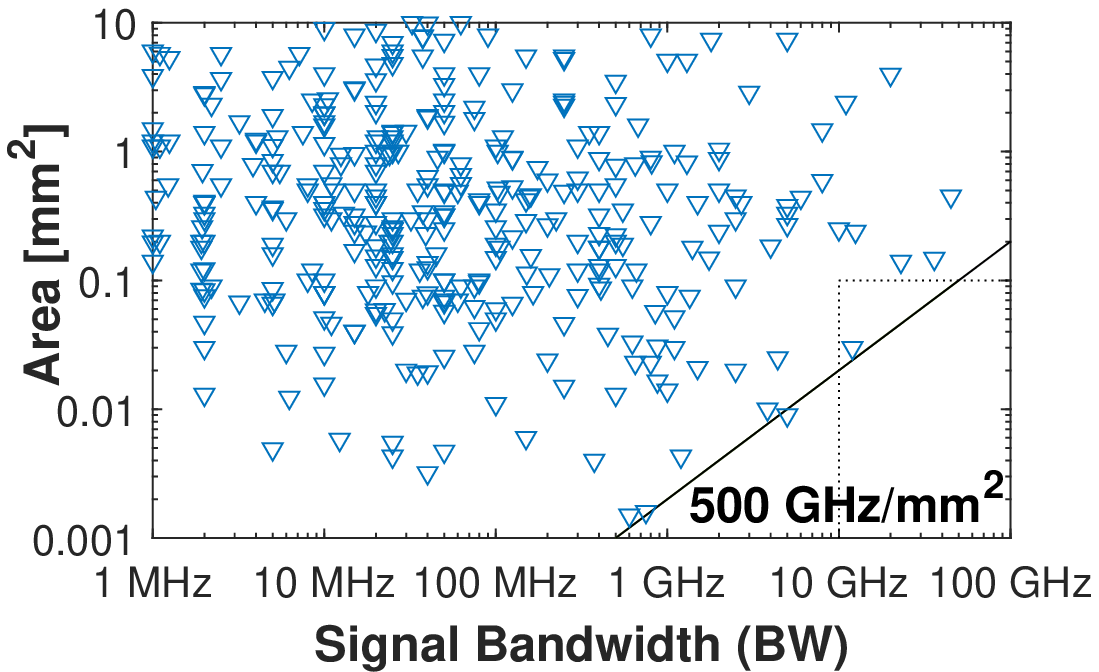}}
\vspace{-0.1cm}
\caption{Snapshot of the historic performance and efficiency of ADCs as functions of the signal bandwidth. Right-side dotted lines delimit \ac{WNoC} targets.}
%\caption{Effective Number of Bits (ENOB) as a function of the signal bandwidth assuming \ac{OOK} modulation.}
%\caption{Energy consumed by the \ac{ADC} for each modulated bit as a function of the signal bandwidth assuming \ac{OOK} modulation. The bottom-right area delimits the \ac{WNoC} target.}
%%\caption{ADC area as a function of the signal bandwidth. The bottom-right area delimits the \ac{WNoC} target.}
\label{fig:current}
\vspace{-0.1cm}
\end{figure*}

\subsection{Performance}
To evaluate performance, we focus on the signal bandwidth $BW$. Since we assume the use of low order modulations, the data rate (in Gb/s) tends to be equal to $BW$ (in GHz). The \ac{ADC} needs to provide sampling at the Nyquist rate $f_{snyq}$ at least, or at higher sampling rates $f_{s}$ for a given oversampling ratio $OSR$, so that 
\begin{equation}
BW = \frac{f_{snyq}}{2} = \frac{f_{s}}{2\cdot OSR}.
\end{equation}

Figure \ref{fig:perf} shows a reinterpretation of the conventional aperture graph, plotting \ac{ENOB} as a function of the signal bandwidth in our case. It is observed that recent \acp{ADC} achieve the required 10 GHz, with the greatest bandwidth being 45 GHz with an impressive Nyquist rate of 90 GS/s \cite{Kull2014}. However, this and subsequent designs at 72 and 64 GS/s could not provide valid \ac{ENOB} measurements at such high bandwidths. Duan \emph{et al.}, instead, are capable of proving 23 GHz of signal bandwidth at Nyquist rate of 46 GS/s with an \ac{ENOB} of around 4 bits \cite{Duan2015}. 

Additionally, Figure \ref{fig:perf} illustrates the upper performance bounds of \acp{ADC} limited by jitter, which helps quantify the maximum admissible noise level at the clocks. Although several designs already pushed the 0.1 ps limit and femtosecond values are possible with photonic alternatives \cite{Khilo2012}, \acp{ADC} for \ac{WNoC} do not need to move past those barriers thanks to their relatively low \ac{ENOB} requirement. 
	
\subsection{Energy Consumption}
To evaluate energy consumption, let us assume that all bits coming from the \ac{ADC} are used for the decoding of symbols modulated at 1 b/s/Hz, as per simplicity requirements at the transceiver. In line with the limited bit depth requirements of \ac{WNoC}, extra \ac{ADC} bits would be wasted. Taking this into consideration, we define the \emph{single-bit energy} $E_{bit}$ as
\begin{equation}
E_{bit} = \frac{P}{BW} = \frac{2P}{f_{snyq}},
\end{equation}
where $P$ is the power consumption at the \ac{ADC}. The results are expressed in pJ/bit and aim to convey a measure of the energy consumed by the \ac{ADC} per each modulated bit, analogous to the bit energy at the transceiver.

Figure \ref{fig:power1} represents the single-bit energy as a function of the signal bandwidth, distinguishing between low and high ENOB designs. A first observation is that \textbf{none} of the reported \acp{ADC} is capable of providing the required efficiency. The closest is the design by Xu \emph{et al.} again, which consumes 23 mW providing 12 GHz of bandwidth and therefore should be at least halved to reach the \ac{WNoC} requirements. Another striking result is that low ENOB designs, which would be theoretically less power-hungry than the high ENOB ones, are very sparse and do not present better efficiencies overall. However, we speculate that the low ENOB requirements of the scenario could help minimize the power consumption.

%%\begin{figure}[!t]
%%\centering
%%\includegraphics[width=\columnwidth]{./figs/energy.eps}
%%%\vspace{-0.35cm}
%%\caption{Energy consumed by the \ac{ADC} for each modulated bit as a function of the signal bandwidth assuming \ac{OOK} modulation. The bottom-right area delimits the \ac{WNoC} target.}
%%\label{fig:power1}
%%%\vspace{-0.3cm}
%%\end{figure}

\subsection{Area Overhead}
In data converters, area has been always accounted for, but never seen as a primary concern. In chip-scale communications, however, the chip real estate is limited and \acp{ADC} should minimize their active area. However, high speeds require either complex circuits or a considerable number stages in time-interleaved architectures, complicating the task of delivering compact yet fast designs.

%%\begin{figure}[!t]
%%\centering
%%\includegraphics[width=\columnwidth]{./figs/area.eps}
%%%\vspace{-0.35cm}
%%\caption{ADC area as a function of the signal bandwidth. The bottom-right area delimits the \ac{WNoC} target.}
%%\label{fig:area1}
%%%\vspace{-0.3cm}
%%\end{figure}

As observed in Figure \ref{fig:area1}, only a 2017 design by Xu \emph{et al.} \cite{Xu2017} is capable of barely meeting the requirements of the \ac{WNoC} scenario thanks to its 0.03 mm\textsuperscript{2} and 12 GHz of bandwidth (Nyquist rate of 24 GS/s) implemented in 28-nm CMOS. A few proposals, including this one, have an ADC area $A$ so that they achieve a \emph{sampling density}% $\delta_{S}$
\begin{equation}
\label{eq:dens}
\delta_{S} = \frac{f_{snyq}}{A}
\end{equation}
of around 500 GHz/mm\textsuperscript{2}. This is about one order of magnitude larger than the bandwidth density of the transceiver (Fig. \ref{fig:transc}). It is expected that as high-speed \acp{ADC} mature and technologies below 32-nm CMOS become widespread, new designs will be able to surpass this barrier and enter the desired target area, as we will see next.

\begin{figure*}[!t]
\centering
\subfigure[Performance ($BW\cdot 2^{ENOB}$)\label{fig:time1}]{\includegraphics[width=0.32\textwidth]{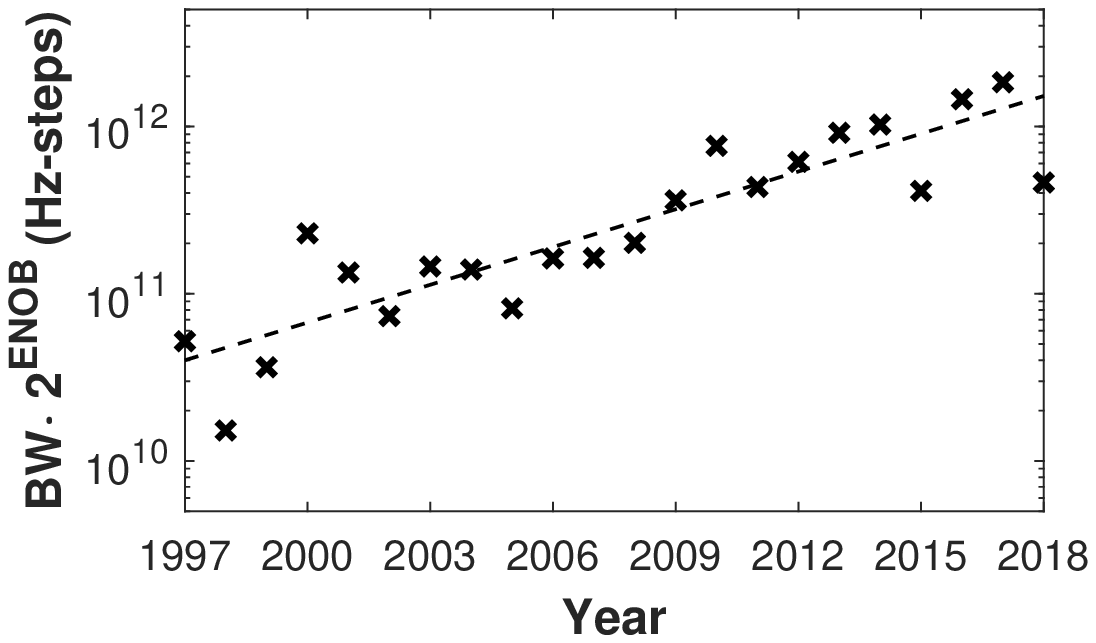}}
\subfigure[Energy ($E_{bit}$)\label{fig:time3}]{\includegraphics[width=0.32\textwidth]{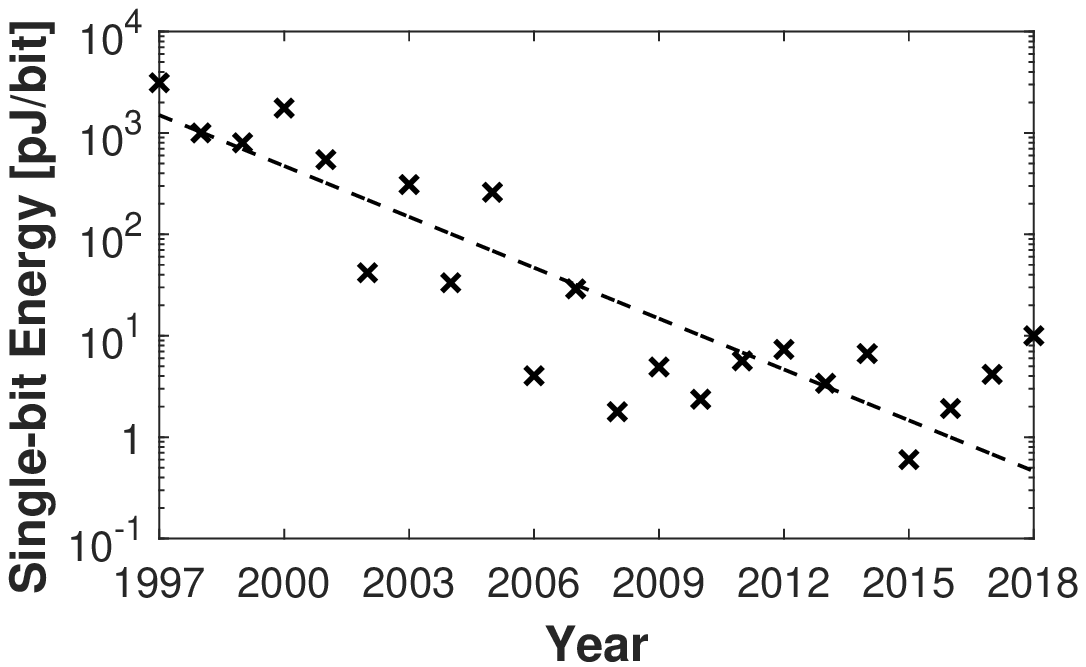}}
\subfigure[Area ($\delta_{S}$)\label{fig:time2}]{\includegraphics[width=0.32\textwidth]{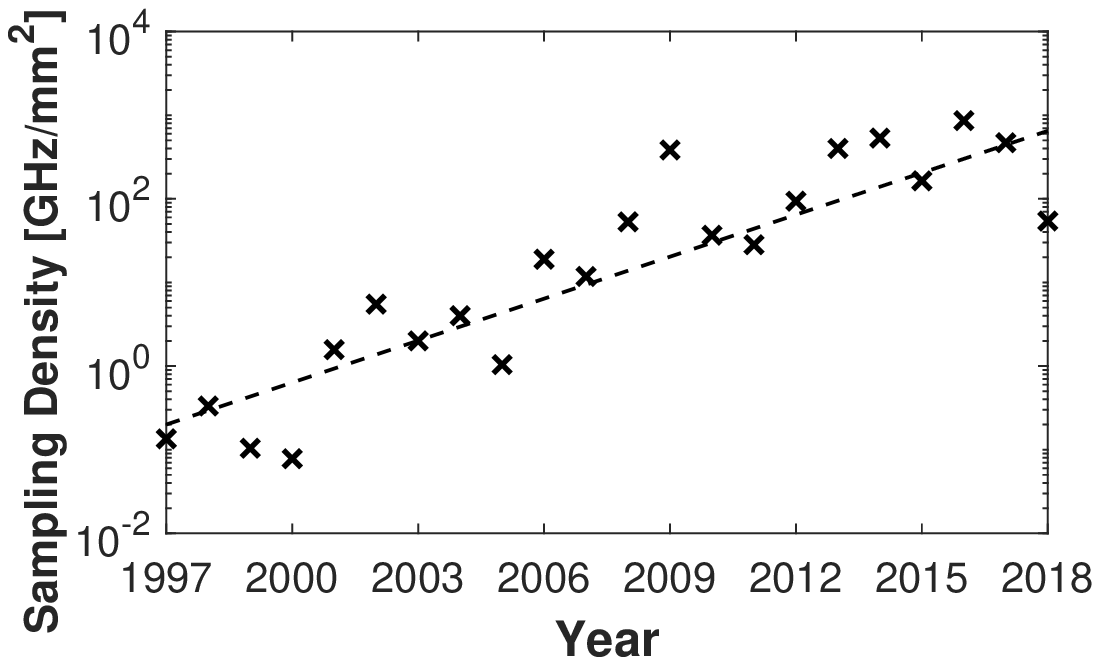}}
\vspace{-0.1cm}
\caption{Yearly evolution of the best ADC designs, together with tendencies identified in \cite{Murmann2015}.}
\label{fig:time}
\vspace{-0.1cm}
\end{figure*}

\section{Discussion: Future Trends}
\label{sec:discussion}
The results obtained in Section \ref{sec:trend} demonstrate that latest \acp{ADC} are at the verge to achieving the performance and efficiency demanded by \ac{WNoC}, but also that improvements are needed especially in terms of energy consumption. Here, we provide an analysis of the on-going scaling trends with the aim to anticipate the feasibility of the data conversion in our target scenario.
%and other ways to simplify the data conversion 

\subsection{Temporal Evolution and Limits}
\textbf{Performance.} Murmann has formulated several scaling trends in several publications by analyzing the designs that push the envelope. For instance, in 2015 he stated that the speed--resolution product ($f_{s}\cdot 2^{ENOB}$) doubles every four years \cite{Murmann2015} and, as observed in Figure \ref{fig:time1}, the trend continues nowadays. Given that the maximum achievable ENOB does not vary much among generations, it is reasonable to affirm that \acp{ADC} with higher speeds will continue to appear in the following years and that 100-Gb/s systems could be a reality at some point.

\textbf{Energy Consumption.} Another trend assessed by Murmann in \cite{Murmann2015} relates to the Schreier's figure of merit, which expresses the energy efficiency of an \ac{ADC} through
\begin{equation}
\label{eq:FOM}
FOM_{S} = SNDR + 10 \log\left(\frac{f_{s}/2}{P}\right),
\end{equation}
where $SNDR = 6.02\cdot ENOB + 1.76$ is the signal-to-noise-distortion ratio. As observed in Figure \ref{fig:timeX}, designs reach a practical limit which has been extended over the years vertically at low frequencies and to the right at high frequencies. In the latter case, which better applies to our scenario, the trend is that the sampling frequency for which we can achieve a given level of $FOM_{S}$ doubles every 1.8 years. 

The evolution of the Schreier's FOM provides the intuition that the energy consumption can be reduced in a similar pace, as we show in Figure \ref{fig:time2}, which plots the best $E_{bit}$ reported each year. Murmann's tendency, also drawn, provides a good approximation of what we can expect in future years. 

Here, it is worth noting that our discussion is far from the fundamental limit on energy consumption. Such limit is given by \emph{the minimum energy required to drive a sampling capacitor using an ideal (Class-B) amplifier} \cite{Murmann2015}
\begin{equation}
\left(\frac{P}{f_{s}}\right)_{min} = 8kT \cdot SNR,
\end{equation}
where $K$ is the Boltzmann constant and $T$ is the temperature, and $SNR\approx SNDR$ is the signal-to-noise ratio. For the low ENOB requirements that we have, this formulation sets the fundamental limit more than 3 orders of magnitude below our most stringent target of 0.1 pJ/bit.

\textbf{Area.} Although no specific area trends have been formulated in the related work, we speculate that the tendency above could also be applied here. To confirm this, we evaluate the sampling density with Equation \eqref{eq:dens} and plot the best value per year. The results shown in Figure \ref{fig:time3} suggests that, indeed, the sampling density also may be doubling every 1.8 years.

\begin{figure*}[!t]
\centering
\subfigure[Energy\label{fig:tech1}]{\includegraphics[width=\columnwidth]{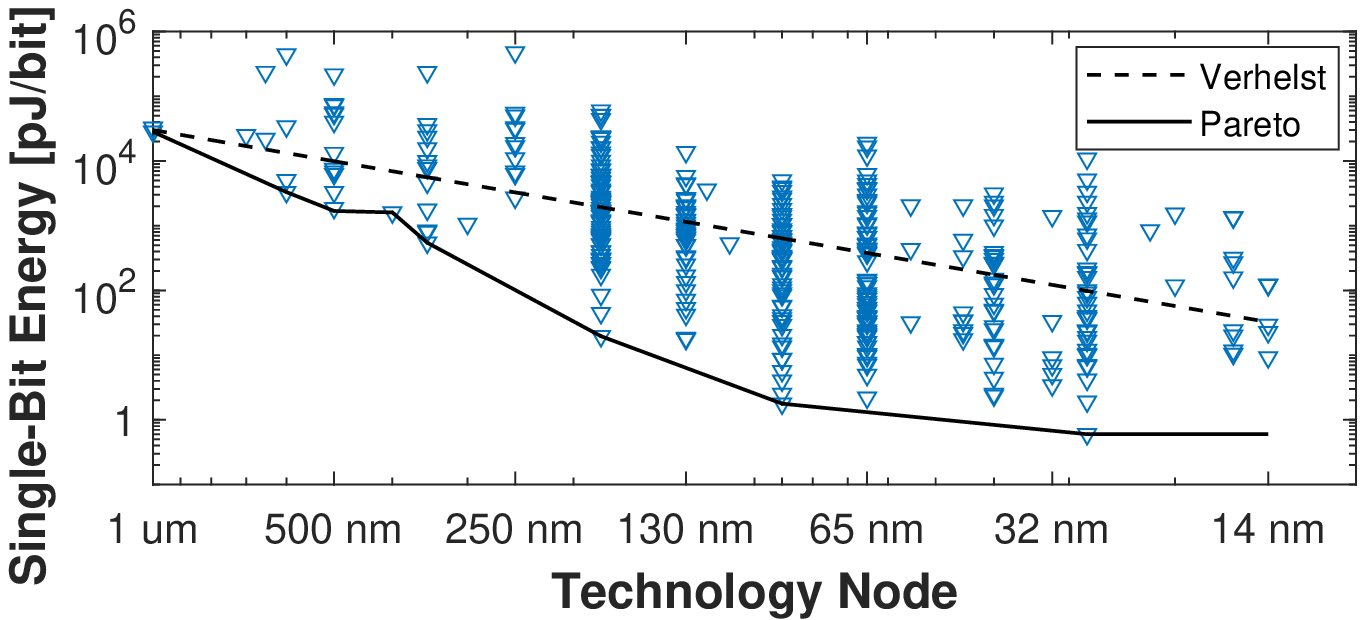}}
\subfigure[Area\label{fig:tech2}]{\includegraphics[width=\columnwidth]{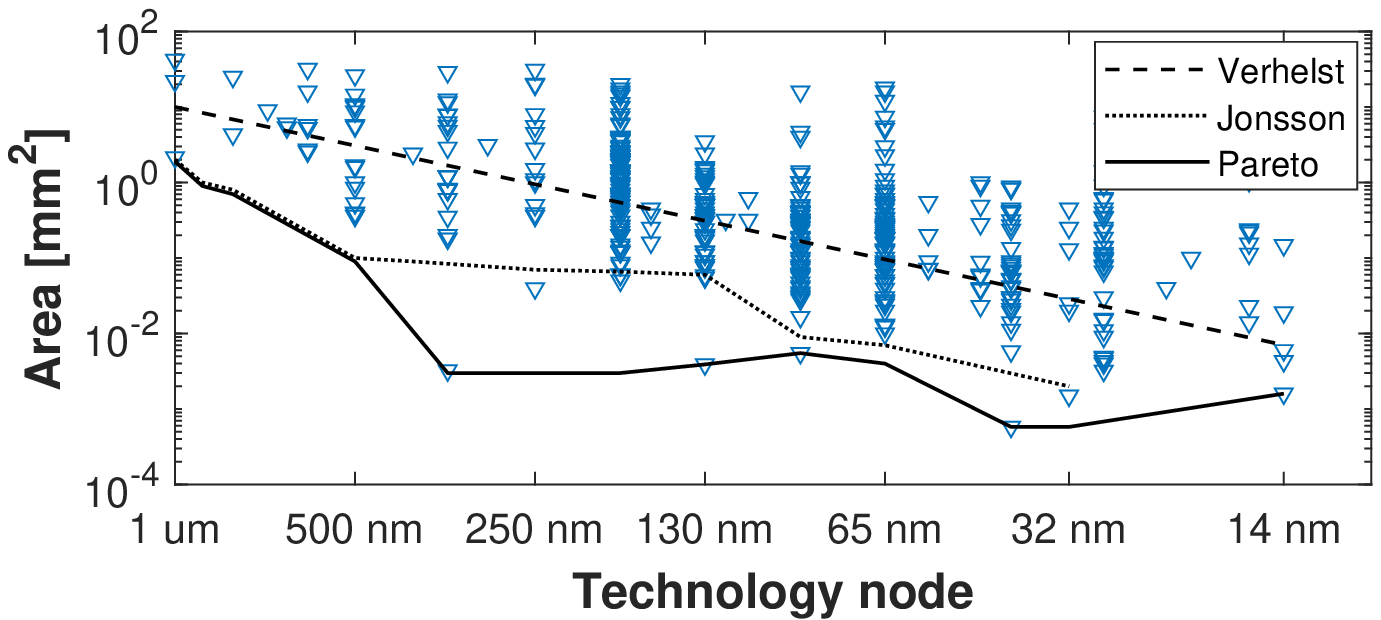}}
\vspace{-0.35cm}
\caption{Downscaling trends of ADCs, including tendencies identified in \cite{Jonsson2011, Verhelst2012} and Pareto-optimal fittings.}
\label{fig:tech}
\vspace{-0.2cm}
\end{figure*}

\subsection{Impact of Technology Downscaling}
The scaling trends demonstrated are product of the downscaling of technology, as well as of circuit optimizations that are realized as new technologies mature. The impact on energy and area is in principle clear, as transistors become smaller, faster, and can be driven with lower voltages. However, side-effects such as leakage or parasitics that appear when pushing the technology to the limit may dilute those advantages. In any case, some works have inspected the impact of technology downscaling. Here, we update the analysis of \cite{Jonsson2011, Verhelst2012} with the 2018 version of Murmann's data. 

In terms of energy, the analysis from \cite{Verhelst2012} predicted an average improvement of $E \sim \lambda^{1.7}$ where $\lambda$ is the technology feature size. Figure \ref{fig:tech1} confirms that thus tendency is a good approximation of average behavior. Note, however, that some \ac{ADC} architectures benefit more from technology scaling. For instance, \ac{SAR} \acp{ADC} obtain up to $\lambda^{2.3}$ as stated in \cite{Verhelst2012} and recently confirmed in \cite{Murmann2016}, turning them into a great choice for high-speed efficient conversion. The work by Xu \emph{et al.}, arguably the best candidate for \ac{WNoC}, is actually a \ac{SAR} \ac{ADC}. 

In terms of area, the works by Jonsson \cite{Jonsson2011} and Verhelst \cite{Verhelst2012} concluded that average behavior scales as $A \sim \lambda^{2}$ and $A \sim \lambda^{1.6}$, respectively. They are both represented in Figure \ref{fig:tech2}, and at first sight the answer by Verhelst seems to better fit the data in average. 

The Pareto optimality analysis seems to imply, that by reaching sub-20nm technologies, we can expect a slight saturation of the area and energy benefits when downscaling. This may be due to tunneling at the transistors and other undesired effects, but also needs confirmation as technology matures and new optimization techniques are worked out.

%\subsection{Advanced Modulations}
%Techniques to reduce burden from Dresden.
%Undersampling.
%
%Other people talking about 1-bit ADCs.

\section{Conclusion and Future Perspectives}%\vspace{-0.2cm}
\label{sec:conclusion}
This paper has provided an estimation of the \ac{ADC} requirements in the \ac{WNoC} scenario. Taking base on low-order modulations, high signal bandwidths (well over 10 GHz) and stringent area and energy limitations (below 1 pJ/bit and 0.1 mm\textsuperscript{2}) are expected. Current \ac{ADC} designs barely meet these demands, but on-going scaling trends suggest that data conversion will not become a bottleneck in \acp{WNoC}. We estimate that high-speed \acp{ADC} at 0.1 pJ/bit and 0.01 mm\textsuperscript{2}, or even below, can be a reality in 5--10 years unless \ac{WNoC}-specific designs are attempted. In such case, one-bit quantization and undersampling techniques could be explored with the aim of softening the technical requirements of \acp{ADC} and paving the way towards the realization of the \ac{WNoC} paradigm.

\section*{Acknowledgments}
The authors thank Alejandro L\'{o}pez-Lao for his technical support and discussions during the development of this work.

% trigger a \newpage just before the given reference
% number - used to balance the columns on the last page
% adjust value as needed - may need to be readjusted if
% the document is modified later
%\IEEEtriggeratref{8}
% The "triggered" command can be changed if desired:
%\IEEEtriggercmd{\enlargethispage{-5in}}

% ---------------------------------------------------
% References
% ---------------------------------------------------

% can use a bibliography generated by BibTeX as a .bbl file
% BibTeX documentation can be easily obtained at:
% http://www.ctan.org/tex-archive/biblio/bibtex/contrib/doc/
% The IEEEtran BibTeX style support page is at:
% http://www.michaelshell.org/tex/ieeetran/bibtex/

% Generated by IEEEtran.bst, version: 1.14 (2015/08/26)

% ---------------------------------------------------
% End of Document
% ---------------------------------------------------

% that's all folks
\end{document}